\documentclass[5pt]{amsart}
\usepackage{graphicx}
\theoremstyle{definition}

\theoremstyle{remark}

\numberwithin{equation}{section}

\begin{document}

\title[]{On $[[n,n-4,3]]_{q}$ Quantum  MDS Codes for odd prime power  $q$ }%
\author{Ruihu Li$^{(1)}$$^{(2)}$,  Zongben Xu$^{(1)}$}%
\thanks{ Ruihu Li is with the College of  Science, Xi'an Jiaotong University,
Shaanxi 710049, People's Republic of China, and College of  Science,
Air Force Engineering University, Xi'an, Shaanxi 710051, People's
Republic of China  (E-mail: liruihu2008@yahoo.com.cn).}

\thanks{Zongben Xu is with the Institute for Information and System
Sciences, Xi'an Jiaotong University, Shaanxi 710049, People's
Republic of China (E-mail: zbxu@mail.xjtu.edu.cn) }

\thanks{ This work is supported by Natural Science Foundation of China  under
Grant No.60573040,  the National Basic Research Program of China
(973 Program) under Grant No.2007CB311002, Natural Science Basic
Research Plan in Shaanxi Province of China under Program No.SJ08A02.
}

\thanks{}%

\begin{abstract} For each odd prime  power $q$, let $4 \leq n\leq
q^{2}+1$. Hermitian self-orthogonal $[n,2,n-1]$ codes  over
$GF(q^{2})$ with dual distance three are constructed by using finite
field theory. Hence, $[[n,n-4,3]]_{q}$
quantum  MDS codes for $4 \leq n\leq q^{2}+1$ are obtained.\\

\end{abstract}
 \maketitle
\section{Introduction}
  The theory of quantum error-correcting codes (QECCs, for short)
was established a decade ago as the primary tool for fighting
decoherence in quantum computers and quantum communication system,
see \cite{shor} and \cite{st}.  The most widely studied class of
quantum codes are binary quantum stabilizer codes. A thorough
discussion on the principles of quantum coding theory was given in
\cite{cal} and \cite{got}  for binary quantum  stabilizer codes. An
appealing aspect of binary quantum codes is that there exist links
to classical coding theory which easy the construction of good
quantum codes. Following \cite{cal} and \cite{got}, many binary
quantum codes are constructed from binary and quaternary classical
codes, see [6-7, 13-18, 23-25].

Almost at the same time of \cite{cal}, some results of binary
quantum stabilizer codes were generalized to the case of non-binary
quantum stabilizer codes, and characterization of non-binary quantum
stabilizer codes over $GF(q)$ in term of classical codes over
$GF(q^{2})$ was also given which generalizes the well-known notation
of additive codes over $GF(4)$ for binary case, see \cite{rain},
\cite{ash} and \cite{ket} and references therein. And, many
non-binary quantum stabilizer codes are constructed from classical
codes over $GF(q)$ or over $GF(q^{2})$, see [2-3,5,8, 10-11].

One central theme in quantum error-correction is the construction of
quantum codes with good parameters. Except the  method of
constructing quantum codes from classical self-orthogonal linear
codes over $GF(q)$  and self-orthogonal additive codes over
$GF(q^{2})$ that given in \cite{cal}, \cite{got},\cite{rain},
\cite{ash} and \cite{ket}, Schlingemann and Werner \cite{sch}
presented another new way to construct quantum stabilizer codes by
finding certain graphs (or matrices) with specific properties. A
number of researchers use these  methods to construct optimal
quantum codes, e.g., codes with largest possible $k$ with fixed $n$
and $d$. Optimal quantum codes that saturating the quantum Singleton
bound received much attention.

{\bf Lemma 1.1: \ \cite{kni} \cite{rain}(quantum Singleton bound)}
An $[[n,k,d]]_{q}$ quantum stabilizer codes satisfies $$k\leq
n-2d+2.$$

A quantum code attains the quantum Singleton bound is called a
quantum maximum distance separable code or {\it quantum MDS code}
for short. It is known that except trivial codes ( codes with $d\leq
2$), there are only two binary  quantum MDS codes, $[[5,1,3]]_{2}$
and $[[6,0,4]]_{2}$, see \cite{cal}. Non-binary  quantum MDS codes
are much complex compared  with the binary case. In the simplest
nontrivial case $d=3$, despite many efforts to construct non-binary
quantum MDS codes, a systematic construction for all lengths has not
been achieved yet, see [2-3,5,8, 10-11]. Known results on non-binary
quantum MDS codes are as follows: \cite{rain} proved the existence
of $[[5,1,3]]_{p}$ for odd prime $p$, \cite{bie} proved the
existence of $[[q^{2}+1,q^{2}-3,3]]_{q}$ quantum MDS codes for prime
power $q$, \cite{gra0} proved the existence of $[[n,n-2d+2,d]]_{q}$
for all $3\leq n\leq q$ and $1\leq d\leq \frac{n}{2}+1$, and
$[[q^{2},q^{2}-2d+2,d]]_{q}$ for $1\leq d\leq q$, all these three
papers used self-orthogonal  codes to construct quantum MDS codes.
 Based on graph states method, \cite{sch} proved the existence of
$[[5,1,3]]_{p}$ for prime  $p\geq 3$, \cite{feng} proved the
existence of $[[6,2,3]]_{p}$ and $[[7,3,3]]_{p}$ for prime $p\geq
3$. With a computer search and  graph states  method, \cite{hu}
constructed four families of quantum MDS codes $[[6,2,3]]_{p}$,
$[[7,3,3]]_{p}$, $[[8,4,3]]_{p}$,  and $[[8,2,4]]_{p}$, for odd
$p\geq 3$.

If $d\geq 3$, \cite{ket} proved that the maximal length $n$ of
 $[[n,k,d]]_{q}$ quantum MDS codes satisfies $n\leq q^{2}+d-2$.
   In this paper, we will use Hermitian self-orthogonal  codes over
$F_{q^{2}}$ to construct $q$-ary quantum MDS codes of distance
three,  where $q=p^{r}$ and $p\geq 3$ is an odd prime. Our main
result of this paper is as follow:

{\bf Theorem 1.1:}  If  $q=p^{r}$ and $p\geq 3$ is an odd prime,
then there are  $[[n,n-4,3]]_{q}$ quantum MDS codes for $4 \leq
n\leq q^{2}+1$.

\section{Preliminaries}

In order to present our main result, we  make some preparation on
quantum codes and finite fields.

Let $GF(q^{2})^{n}$ be the  $n$-dimensional row space over the
finite field $GF(q^{2})$.  For $X=(x_{1},x_{2},...,x_{n})$,
$Y=(y_{1},y_{2},...,y_{n})\in$ $GF(q^{2})^{n}$, the Hermitian
 inner product of $X$ and $Y$ is defined as follow:
$$(X,Y)=x_{1}y_{1}^{q}+x_{2}y_{2}^{q}+...+x_{n}y_{n}^{q}.$$
If $\mathcal{C}$ is an $[n,k]$ linear  code over $GF(q^{2})$, its
dual code $\mathcal{C}^{\perp}$ $=\{X \mid X \in GF(q^{2})^{n}, (X,
Y)=0 \  \hbox{for any }  Y \in GF(q^{2})^{n}  \}.$ $\mathcal{C}$ is
self-orthogonal if $\mathcal{C}\subseteq$ $\mathcal{C}^{\perp}$,
 and self-dual if $\mathcal{C}=$ $\mathcal{C}^{\perp}$.

 \cite{gra0} gave the following theorem of constructing quantum
codes from self-orthogonal codes over $GF(q^{2})$.

{\bf Theorem 2.1: }  If $\mathcal{C}$ is an $[n,k]$ linear  code
over $GF(q^{2})$ such that $\mathcal{C}^{\perp}\subseteq$
$\mathcal{C}$, and $d=min\{wt(v):v\in \mathcal{C}\setminus
\mathcal{C}^{\perp}\}$, then there exists $[[n,2k-n, d]]_{q}$
quantum code.

In the construction of self-orthogonal codes, we also need the
following results on finite fields.

{\bf Lemma 2.1:} If $\alpha$ is a primitive element of $GF(q^{2})$,
for each non zero element $\beta$ of $GF(q)$, there are $q+1$
elements $\alpha^{i}$ of $GF(q^{2})$ such that $(\alpha^{i})^{q+1}$
$=\beta$.

{\bf Proof:} Suppose $(\alpha)^{q+1}=\gamma$, then  $\gamma$ is a
primitive element of $GF(q)$. Let $\beta$ $=\gamma^{i}$, $0\leq
i\leq q-2$. Then $(\alpha^{i+(q-1)j})^{q+1}=\beta$ for $0\leq j\leq
q$, thus the Lemma holds.

{\bf Lemma 2.2:} If $\alpha$ is a primitive element of $GF(q^{2})$,
then
$1+(\alpha)^{q+1}+(\alpha^{2})^{q+1}+...+(\alpha^{q^{2}-2})^{q+1}$
$=0$.

{\bf Proof:} Suppose $(\alpha)^{q+1}=\gamma$, then
$1+(\alpha)^{q+1}+(\alpha^{2})^{q+1}+...+(\alpha^{q^{2}-2})^{q+1}$
$=(q+1)(1+\gamma+\gamma^{2}+...+\gamma^{q-2})=0$

To simplify statement of the following two section, we divide the
non-zero elements of $GF(q^{2})$ into two subsets, say $A$ and $B$.
Let $A=\{ 1, -1, \alpha^{\frac{q-1}{2}}, -\alpha^{\frac{q-1}{2}},
\alpha^{\frac{q-1}{2}}+1,
  -\alpha^{\frac{q-1}{2}}-1 \}$, $B=\{ x_{1}, -x_{1}, ...,
x_{k}, -x_{k} \}$ $=GF(q^{2})\setminus (A\cup \{0\})$, where
$2k=q^{2}-7$.

According to  Lemma 2.2, one can deduce  that

{\bf Lemma 2.3:} Suppose $A$ and $B$ be defined as above, then
$2\Sigma_{i=1}^{k}(x_{i})^{q+1}+2(\alpha^{\frac{q-1}{2}}+1)^{q+1}=0$.

 \section{ $[[n,n-4,3]]_{q}$ for $q=3^{r}$ }
In this section, we will prove Theorem 1.1 holds for $q=3^{r}$.

First we discuss the construction of $[[n,n-4,3]]_{3}$ quantum code.

 Let $GF(3)=\{0,1,2\}=\{0,1,-1\}$ be the Galois field with three
elements. then $f(x)=x^{2}+x+2$ is irreducible over $GF(3)$. Using
$f(x)$, one can construct the Galois field $GF(9)$ with nine
elements as $GF(9)=\{0,1,2, \alpha, \alpha+1, \alpha+2, 2\alpha,
2\alpha+1, 2\alpha+2\}$, where $\alpha$ is a root of
$f(x)=x^{2}+x+2$. It is easy to check that $\alpha$ is a primitive
element of $GF(9)$, $\alpha^{2}=2\alpha+1$, $\alpha^{3}=2\alpha+2$,
$\alpha^{4}=2$, $\alpha^{5}=2\alpha$, $\alpha^{6}=\alpha+2$, and
 $\alpha^{7}=\alpha+1$. It is obvious that $\alpha^{4+i}=-\alpha^{i}$
 for $0 \leq i \leq 7 $.

 Construct
$$
H _{2,4}= \left(
\begin{array}{cccccccccc}
1&1&1&0\\
0&1&-1&1\\
\end{array}
\right),
 H _{2,5}= \left(
\begin{array}{cccccccccc}
1&1&\alpha&\alpha&0\\
0&1&\alpha^{2}&\alpha^{3}&\alpha\\
\end{array}
\right),
$$
$$
H _{2,6}= \left(
\begin{array}{cccccccccc}
1&\alpha&\alpha&\alpha&\alpha&0\\
0&1&\alpha^{2}&\alpha^{4}&\alpha^{6}&\alpha\\
\end{array}
\right),
 H _{2,7}= \left(
\begin{array}{cccccccccc}
1&1&1&1&1&1&0\\
0&1&\alpha^{1}&\alpha^{2}&\alpha^{5}&\alpha^{7}&1\\
\end{array}
\right),
$$
$$
H _{2,8}= \left(
\begin{array}{cccccccccc}
1&1&1&1&1&\alpha&\alpha&0\\
0&1&2&\alpha^{1}&\alpha^{5}&1&2&1\\
\end{array}
\right),
 H _{2,9}= \left(
\begin{array}{cccccccccc}
1&1&1&...&1\\
0&1&\alpha^{1}&...&\alpha^{7}\\
\end{array}
\right),
$$
$$
 H _{2,10}= \left(
\begin{array}{cccccccccc}
1&1&1&1&1&1&\alpha&\alpha&\alpha&0\\
0&1&\alpha&\alpha^{4}&\alpha^{6}&\alpha^{7}&\alpha^{3}&\alpha^{4}&\alpha^{6}&1\\
\end{array}
\right),
$$
It is easy to check that: For $4\leq n\leq 10$, the code
$\mathcal{C}_{2,n}$ generated by $H _{2,n}$ is a  self-orthogonal
code over $GF(9)$ and $d^{\perp}=3$, hence there is an
$[[n,n-4,3]]_{3}$ quantum MDS codes.

Second, we discuss the construction of $[[n,n-4,3]]_{q}$ quantum
code for $q=3^{r}\geq 9$. To achieve this, we consider four cases
separately.

{\bf Case 3.1} Let $4\leq n \leq q^{2}-4$ and $ n \equiv 0( mod \
2)$.

 Let $ n - 4=2k_{1}$, $ u=2\Sigma_{i=1}^{k_{1}}(x_{i})^{q+1}$.
Choose $\gamma,\delta,\epsilon \in GF(q^{2})$ such that
$\delta^{q+1}\in GF(q)\setminus GF(3)$  and $2\delta^{q+1}+u\neq 0$,
 $\gamma^{q+1}=-(n-4+2\delta^{q+1})$,
 $\epsilon^{q+1}=-(u+2\delta^{q+1})$.
Construct
 $$
A_{2,n} = \left(
\begin{array}{cccccccccc}
\gamma&1&1&\cdots&1&1&\delta&\delta&0\\
0&x_{1}&-x_{1}&\cdots&x_{k_{1}}&-x_{k_{1}}&\delta&-\delta&\epsilon\\
\end{array}
\right).
$$

{\bf Case 3.2} Let $4\leq n \leq q^{2}-4$ and $ n \equiv 1( mod \
2)$.

 Let $ n - 5=2k_{1}$, $ u=2\Sigma_{i=1}^{k_{1}}(x_{i})^{q+1}$.
Choose $\gamma,\delta,\epsilon \in GF(q^{2})$, such that
$\delta^{q+1}\in GF(q)\setminus GF(3)$ and
$u+\delta^{q+1}(\alpha^{q-1}+1)^{q+1}\neq 0$,
 $\gamma^{q+1}=-(n-5+\delta^{q+1})$,  $\epsilon^{q+1}=-(u+\delta^{q+1}(\alpha^{q-1}+1)^{q+1})$. Construct
 $$
A_{2,n} = \left(
\begin{array}{ccccccccccc}
\gamma&1&1&\cdots&1&1&\delta \alpha^{\frac{q-1}{2}}&\delta&\delta&0\\
0&x_{1}&-x_{1}&\cdots&x_{k_{1}}&-x_{k_{1}}&\delta \alpha^{\frac{q-1}{2}}
&-\delta \alpha^{q-1}&\delta(\alpha^{q-1}+1)&\epsilon\\
\end{array}
\right).
$$

{\bf Case 3.3} Let $q^{2}-3\leq n \leq q^{2}-1$.

 {\bf Subcase 3.3.1}
If $ n = q^{2}-1$, choose $\gamma,\delta, \epsilon \in GF(q^{2})$
such that   $\delta^{q+1}\in GF(q)\setminus GF(3)$ and
$1-\delta^{q+1}-(\alpha^{q-1}+1)^{q+1}\neq
0$,   $\gamma^{p+1}=5-2\delta^{q+1}$,  $\epsilon^{q+1}=2(\delta^{q+1}+(\alpha^{\frac{q-1}{2}}+1)^{q+1}-1)$.
  Construct
$$
A_{2,n} = \left(
\begin{array}{ccccccccccccc}
\gamma&1&1&\cdots&1&1&1&1&\delta&\delta&0\\
0&x_{1}&-x_{1}&\cdots&x_{k}&-x_{k}&1&-1&\delta\alpha^{\frac{q-1}{2}}&-\delta\alpha^{\frac{q-1}{2}}
&\epsilon\\
\end{array}
\right).
$$
{\bf Subcase 3.3.2}  If $ n = q^{2}-2$, choose $ \epsilon \in
GF(q^{2})$  such that
     $\epsilon=\alpha^{\frac{q-1}{2}}+1$, and construct
$$
A_{2,n} = \left(
\begin{array}{ccccccccccccc}
1&1&1&\cdots&1&1&1&1&1&0\\
0&x_{1}&-x_{1}&\cdots&x_{k}&-x_{k}&1&\alpha^{\frac{q-1}{2}}&-\alpha^{\frac{q-1}{2}}-1
&\epsilon\\
\end{array}
\right).
$$
{\bf Subcase 3.3.3} If $ n = q^{2}-3$, choose $\gamma,\delta,
\epsilon \in GF(q^{2})$
 such that $\delta^{q+1}\in
GF(q)\setminus GF(3)$,   $\gamma^{p+1}=1-2\delta^{q+1}$,
  $\epsilon^{q+1}=(2-2\delta^{q+1})(\alpha^{\frac{q-1}{2}}+1)^{q+1}$. Construct
$$
A_{2,n} = \left(
\begin{array}{ccccccccccccc}
\gamma&1&1&\cdots&1&1&\delta&\delta&0\\
0&x_{1}&-x_{1}&\cdots&x_{k}&-x_{k}&\delta(\alpha^{\frac{p-1}{2}}+1)&-\delta(\alpha^{\frac{p-1}{2}}+1)
&\epsilon\\
\end{array}
\right).
$$

{\bf Case 3.4} Let $q^{2}\leq n \leq q^{2}+1$.

 If $n=q^{2}$,
construct
$$
A_{2,n} = \left(
\begin{array}{ccccccccccccc}
1&1&1&\cdots&1&1\\
0&1&\alpha&\cdots&\alpha^{q^{2}-3}&\alpha^{q^{2}-2}\\
\end{array}
\right).
$$

If $n=q^{2}+1$, let $a=\frac{q-1}{2}$. Choose $\delta, \epsilon \in
GF(q^{2})$, such that
   $\delta^{q+1}\in GF(q)\setminus GF(3)$,
$\epsilon^{q+1}=(1-\delta^{q+1})(\alpha^{\frac{q-1}{2}}+1)^{q+1}$,
and construct
$$
A_{2,n} = \left(
\begin{array}{ccccccccccccc}
1&1&1&\cdots&1&1&1&1&1&\delta&\delta&\delta&0\\
0&x_{1}&-x_{1}&\cdots&x_{k}&-x_{k}&1&\alpha^{a}&-(\alpha^{a}+1)&-\delta&-\delta\alpha^{a}&\delta(\alpha^{a}+1)
&\epsilon\\
\end{array}
\right).
$$
It is easy to check that: In the above four cases, the code
 generated by $A_{2,n}$ is an $[n,2,n-1]$ self-orthogonal code over
$GF(q^{2})$, and its dual distance is 3. Hence, there are
$[[n,n-4,3]]_{q}$ quantum MDS codes for  $4\leq n\leq q^{2}+1$,
 where $q=3^{r}$ .

Summarizing the above discussion,  Theorem 1.1 holds for $q=3^{r}.$

\section{$[[n,n-4,3]]_{q}$ for $q=p^{r}$ and prime $p\geq 5$}
In this section, we will prove Theorem 1.1 holds for $q=p^{r}$, and
 we always assume that  $p\geq 5$ is a prime and
$\alpha$ is a primitive element of $GF(q^{2})$. To give the
construction of quantum $[[n,n-4,3]]_{q}$ codes, we consider four
cases separately.

{\bf Case 4.1} Let $4\leq n \leq q^{2}-4$ and $ n =mp+2$.\\

{\bf Subcase 4.1.1}  If $ n \equiv 0( mod \ 2)$, let $ n - 4=2k_{1}$. \\

If $ v=2\Sigma_{i=1}^{k_{1}}(x_{i})^{q+1}+4\neq 0$, choose
$\gamma,\delta,\epsilon \in GF(q^{2})$ such that
 $\gamma^{q+1}=-2$, $\delta^{q+1}=2$, $\epsilon^{q+1}=-v$. Construct\\
$$
A_{2,n} = \left(
\begin{array}{cccccccccc}
\gamma&1&1&\cdots&1&1&\delta&\delta&0\\
0&x_{1}&-x_{1}&\cdots&x_{k_{1}}&-x_{k_{1}}&\delta&-\delta&\epsilon\\
\end{array}
\right).
$$

If $ v=2\Sigma_{i=1}^{k_{1}}(x_{i})^{p+1}+4= 0$, choose
$\gamma,\delta,\epsilon \in GF(q^{2})$, such that
 $\gamma^{q+1}=-2$, $\delta^{q+1}=2$, $\epsilon^{q+1}=8$, and construct\\
$$
A_{2,n} = \left(
\begin{array}{cccccccccc}
\gamma&1&1&\cdots&1&1&\delta&\delta&0\\
0&x_{1}&-x_{1}&\cdots&x_{k_{1}}&-x_{k_{1}}&\delta\alpha^{\frac{q-1}{2}}&-\delta\alpha^{\frac{q-1}{2}}&\epsilon\\
\end{array}
\right).
$$

{\bf Subcase 4.1.2} If $ n \equiv 1( mod \ 2)$, let $ n -5=2k_{1}$. \\

If $
v'=2\Sigma_{i=1}^{k_{1}}(x_{i})^{q+1}+2(\alpha^{\frac{q-1}{2}}+1)^{q+1}\neq
0 $, choose $\gamma,\delta,\epsilon \in GF(q^{2})$ such that
 $\gamma^{q+1}=-3$, $\delta^{q+1}=2$ and $\epsilon^{q+1}=-v'$. Construct\\
$$
A_{2,n} = \left(
\begin{array}{ccccccccccc}
\gamma&1&1&\cdots&1&1&\delta&\delta&\delta&0\\
0&x_{1}&-x_{1}&\cdots&x_{k_{1}}&-x_{k_{1}}&\delta&\delta\alpha^{\frac{q-1}{2}}&-\delta(\alpha^{\frac{q-1}{2}}+1)&\epsilon\\
\end{array}
\right).
$$

If $
v'=2\Sigma_{i=1}^{k_{1}}(x_{i})^{q+1}+2(\alpha^{\frac{q-1}{2}}+1)^{p+1}=
0$, choose $\gamma,\delta,\epsilon \in GF(q^{2})$ such that
  $\gamma^{q+1}=-6$, $\delta^{q+1}=3$ and $\epsilon^{q+1}=-(\alpha^{\frac{q-1}{2}}+1)^{q+1}$. Construct\\
$$
A_{2,n} = \left(
\begin{array}{ccccccccccc}
\gamma&1&1&\cdots&1&1&\delta&\delta&\delta&0\\
0&x_{1}&-x_{1}&\cdots&x_{k_{1}}&-x_{k_{1}}&\delta&\delta\alpha^{\frac{q-1}{2}}&-\delta(\alpha^{\frac{q-1}{2}}+1)
&\epsilon\\
\end{array}
\right).
$$

{\bf Case 4. 2} Let $4\leq n \leq q^{2}-4$ and $ n-2\neq 0(mod \
p)$.

{\bf Subcase 4.2.1}  If $ n \equiv 0( mod \ 2)$, let $ n -
4=2k_{1}$.

If $ w=2\Sigma_{i=1}^{k_{1}}(x_{i})^{q+1}+2\neq 0$, choose
$\gamma,\epsilon \in GF(q^{2})$ such that
 $\gamma^{q+1}=-n+2$ and  $\epsilon^{q+1}=-w$. Construct\\
$$
A_{2,n} = \left(
\begin{array}{cccccccccc}
\gamma&1&1&\cdots&1&1&1&1&0\\
0&x_{1}&-x_{1}&\cdots&x_{k_{1}}&-x_{k_{1}}&1&-1&\epsilon\\
\end{array}
\right).
$$

If $ w=2\Sigma_{i=1}^{k_{1}}(x_{i})^{q+1}+2= 0$, choose
$\gamma,\epsilon \in GF(q^{2})$ such that
 $\gamma^{q+1}=-n+2$ and  $\epsilon^{q+1}=4$. Construct\\
$$
A_{2,n} = \left(
\begin{array}{cccccccccc}
\gamma&1&1&\cdots&1&1&1&1&0\\
0&x_{1}&-x_{1}&\cdots&x_{k_{1}}&-x_{k_{1}}&\alpha^{\frac{q-1}{2}}&-\alpha^{\frac{q-1}{2}}&\epsilon\\
\end{array}
\right).
$$

{\bf Subcase 4.2.2} If $ n \equiv 1( mod \ 2)$, let $ n -5=2k_{1}$.

If $
w'=2\Sigma_{i=1}^{k_{1}}(x_{i})^{q+1}+(\alpha^{\frac{q-1}{2}}+1)^{q+1}\neq
0 $, choose $\gamma,\epsilon \in GF(q^{2})$ such that
 $\gamma^{q+1}=-n+2$,  $\epsilon^{q+1}=-w_{1}$. Construct\\
$$
A_{2,n} = \left(
\begin{array}{ccccccccccc}
\gamma&1&1&\cdots&1&1&1&1&1&0\\
0&x_{1}&-x_{1}&\cdots&x_{k_{1}}&-x_{k_{1}}&1&\alpha^{\frac{q-1}{2}}&-\alpha^{\frac{q-1}{2}}-1&\epsilon\\
\end{array}
\right).
$$

If $
w'=2\Sigma_{i=1}^{k_{1}}(x_{i})^{q+1}+(\alpha^{\frac{q-1}{2}}+1)^{q+1}=
0$ and $n+1\neq 0(mod \ p)$, choose $\gamma,\delta,\epsilon \in
GF(q^{2})$ such that
  $\gamma^{q+1}=-n-1$, $\delta^{q+1}=2$ and $\epsilon^{q+1}=-(\alpha^{\frac{q-1}{2}}+1)^{q+1}$. Construct
$$
A_{2,n} = \left(
\begin{array}{ccccccccccc}
\gamma&1&1&\cdots&1&1&\delta&\delta&\delta&0\\
0&x_{1}&-x_{1}&\cdots&x_{k_{1}}&-x_{k_{1}}&\delta&\delta\alpha^{\frac{q-1}{2}}&-\delta\alpha^{\frac{q-1}{2}}-\delta
&\epsilon\\
\end{array}
\right).
$$

If $
w'=2\Sigma_{i=1}^{k_{1}}(x_{i})^{q+1}+(\alpha^{\frac{q-1}{2}}+1)^{q+1}=
0$ and $n+1\equiv 0(mod \ p)$, choose $\gamma,\delta,\epsilon \in
GF(q^{2})$ such that
  $\gamma^{q+1}=-n-4$, $\delta^{q+1}=3$ and $\epsilon^{q+1}=-2(\alpha^{\frac{q-1}{2}}+1)^{q+1}$. Construct\\
$$
A_{2,n} = \left(
\begin{array}{ccccccccccc}
\gamma&1&1&\cdots&1&1&\delta&\delta&\delta&0\\
0&x_{1}&-x_{1}&\cdots&x_{k_{1}}&-x_{k_{1}}&\delta&\delta\alpha^{\frac{q-1}{2}}&-\delta\alpha^{\frac{q-1}{2}}-\delta
&\epsilon\\
\end{array}
\right).
$$

{\bf Case 4.3} Let $q^{2}-3\leq n\leq q^{2}-1$.

{\bf Subcase 4.3.1}  If $ n = q^{2}-1$, choose $\gamma,\epsilon \in
GF(q^{2})$ such that
  $\gamma^{q+1}=3$,  $\epsilon^{q+1}=2(\alpha^{\frac{q-1}{2}}+1)^{q+1}$, and construct
$$
A_{2,n} = \left(
\begin{array}{ccccccccccccc}
\gamma&1&1&\cdots&1&1&1&1&1&1&0\\
0&x_{1}&-x_{1}&\cdots&x_{k}&-x_{k}&1&-1&\alpha^{\frac{q-1}{2}}&-\alpha^{\frac{q-1}{2}}
&\epsilon\\
\end{array}
\right).
$$

{\bf Subcase 4.3.2}  If $ n = q^{2}-2$, choose $\gamma,\epsilon \in
GF(q^{2})$ such that
  $\gamma^{q+1}=4$,  $\epsilon^{q+1}=(\alpha^{\frac{q-1}{2}}+1)^{q+1}$. Construct
$$
A_{2,n} = \left(
\begin{array}{ccccccccccccc}
\gamma&1&1&\cdots&1&1&1&1&1&0\\
0&x_{1}&-x_{1}&\cdots&x_{k}&-x_{k}&1&\alpha^{\frac{q-1}{2}}&-\alpha^{\frac{q-1}{2}}-1
&\epsilon\\
\end{array}
\right).
$$

{\bf Subcase 4.3.3} Let $ n = q^{2}-3$. Choose $\gamma,\delta,
\epsilon \in GF(q^{2})$ such that
  $\gamma^{q+1}=3$,  $\delta^{q+1}=2$, $\epsilon^{q+1}=-2(\alpha^{\frac{q-1}{2}}+1)^{q+1}$, and construct\\
$$
A_{2,n} = \left(
\begin{array}{ccccccccccccc}
\gamma&1&1&\cdots&1&1&\delta&\delta&0\\
0&x_{1}&-x_{1}&\cdots&x_{k}&-x_{k}&\delta(\alpha^{\frac{q-1}{2}}+1)&-\delta(\alpha^{\frac{q-1}{2}}+1)
&\epsilon\\
\end{array}
\right).
$$

{\bf Case 4.4} Let $q^{2}\leq n\leq q^{2}+1$.

 If $ n = q^{2}$,  construct
$$
A_{2,n} = \left(
\begin{array}{ccccccccccccc}
1&1&1&1&\cdots&1&1\\
0&1&\alpha&\alpha^{2}&\cdots&\alpha^{p^{2}-3}&\alpha^{p^{2}-2}
\\
\end{array}
\right).
$$

 If $n=q^{2}+1$, let $a=\frac{q-1}{2}$. Choose $\gamma, \delta,
\epsilon \in GF(q^{2})$ such that  $\gamma^{q+1}=-2$,
   $\delta^{q+1}=2$ and
$\epsilon^{q+1}=-(\alpha^{\frac{q-1}{2}}+1)^{q+1}$. Construct
$$
A_{2,n} = \left(
\begin{array}{ccccccccccccc}
\gamma&1&1&\cdots&1&1&1&1&1&\delta&\delta&\delta&0\\
0&x_{1}&-x_{1}&\cdots&x_{k}&-x_{k}&1&\alpha^{a}&-(\alpha^{a}+1)&-\delta&-\delta\alpha^{a}&\delta(\alpha^{a}+1)
&\epsilon\\
\end{array}
\right).
$$

It is easy to check that: In the above four cases, the code
 generated by $A_{2,n}$ is an $[n,2,n-1]$ self-orthogonal
code over $GF(q^{2})$ with dual distance is 3. Hence, there are
$[[n,n-4,3]]_{q}$ quantum MDS codes for $4\leq n\leq q^{2}+1$,
 where $q=p^{r}$ and $p\geq 5$ .

Summarizing the above discussion,  Theorem 1.1 holds for $q=p^{r}$
and $p\geq 5$ is a prime.

\section{Concluding Remarks}
For each odd prime power $q$, we have constructed an
$[[n,n-4,3]]_{q}$ quantum MDS codes for $4 \leq n\leq q^{2}+1$. If
$q=2^{r}\geq 4$, are there $[[n,n-4,3]]_{q}$ quantum MDS codes for
$4 \leq n\leq q^{2}+1$? If such quantum MDS codes exist, how to
construct them? These need further study.

\bibliographystyle{amsplain}

\end{document}